\documentclass{ws-p8-50x6-00}

\begin{document}


\thispagestyle{empty}

\begin{flushright}
hep-ph/0006298
\end{flushright}

\vspace{2.0cm}

\begin{center}
\Large\bf Inclusive heavy hadron decays and light-cone dynamics
\vspace*{0.3truecm}
\end{center}

\vspace{1.8cm}

\begin{center}
\large Changhao Jin\\
{\sl School of Physics, University of Melbourne
\\
Victoria 3010, Australia\\[3pt]
E-mail: {\tt jin@physics.unimelb.edu.au
         }}
\end{center}

\vspace{1.5cm}

\begin{center}
{\bf Abstract}\\[0.3cm]
\parbox{13cm}
{
The governing role of light-cone dynamics in inclusive heavy hadron decay 
processes is demonstrated. Nonperturbative QCD effects on the processes
can be systematically calculated using light-cone expansion and
heavy quark effective theory. The applications of the light-cone approach
to studying electroweak and strong interactions and hadron structure with 
semileptonic and radiative decays of beauty hadrons are briefly reviewed.
}
\end{center}

\vspace{2.5cm}

\begin{center}
{\sl To appear in Proceedings of  
the Workshop on Light-Cone QCD and Nonperturbative Hadron Physics\\ 
Adelaide, Australia, December 13-21, 1999}
\end{center}

\newpage
\setcounter{page}{1}
\thispagestyle{empty}


\title{Inclusive Heavy Hadron Decays and Light-Cone Dynamics}

\author{Changhao Jin}

\address{School of Physics, University of Melbourne, Victoria 3010,
Australia\\E-mail: jin@physics.unimelb.edu.au}


\maketitle

\abstracts{
The governing role of light-cone dynamics in inclusive heavy hadron decay 
processes is demonstrated. Nonperturbative QCD effects on the processes
can be systematically calculated using light-cone expansion and
heavy quark effective theory. The applications of the light-cone approach
to studying electroweak and strong interactions and hadron structure with 
semileptonic and radiative decays of beauty hadrons are briefly reviewed.
}

\section{Introduction}
The beauty hadron is the heaviest hadron containing a single heavy
quark but lives for a long time, while the top quark is too heavy to build 
hadrons and has a very short lifetime. This
confers a special role to beauty hadrons in heavy hadron physics: The
longevity of the beauty hadron makes it phenomenologically interesting, and 
the heaviness of the beauty hadron ensures theoretical reliability.
 
Because of the long lifetime of the beauty hadron, many interesting phenomena
have chance to be observed. We discuss here semileptonic or radiative 
inclusive decays of $B$ mesons.
The discussion can be easily extended to inclusive decays
of other heavy hadrons. Semileptonic or radiative inclusive decays of $B$ 
mesons are interesting because these decays are our main tool to determine the
fundamental Cabibbo-Kobayashi-Maskawa (CKM) matrix elements, to probe hadron
structure and strong interactions, to test the standard model, and to search 
for new physics beyond the standard model.

The principal topic in the theoretical description of inclusive decays of $B$ 
mesons is the calculation of nonperturbative QCD contributions. 
There are essentially two components
of nonperturbative QCD effects: one on dynamics --- it 
is responsible for the confinement of
quarks inside a hadron, the other on kinematics --- it transforms parton 
kinematics into hadron kinematics. The past decade
saw great progress in the QCD treatment of inclusive decays of heavy 
hadrons, including the construction of the heavy quark effective 
theory \cite{hqet} and the developments of 
the heavy quark expansion approach \cite{chay} and the light-cone 
approach \cite{jp1,jin1,jp2,jin2,eur}. In this talk I describe
the light-cone approach and present a brief review of some of its applications.
A comparison of the light-cone approach and the heavy quark expansion 
approach can be found in Refs.~\cite{prd,cern}.

The light-cone approach uses the methods of light-cone expansion and heavy 
quark effective theory to address both dynamic and kinematic 
effects of nonperturbative QCD. This QCD approach has led to an improvement 
over the naive parton model \cite{parton} for inclusive $B$ decays, 
so that model-independent predictions and a control over
theoretical uncertainties become possible. 

\section{Light-Cone Expansion}
All nonperturbative QCD physics for 
inclusive semileptonic $B$ decays are incorporated in the hadronic tensor:
\begin{equation}
W_{\mu\nu}= -\frac{1}{2\pi}\int d^4y\,\, e^{iq\cdot y}
\langle B\left|[j_{\mu}(y),j^{\dagger}_{\nu}(0)]\right|B\rangle ,
\label{eq:comm2}
\end{equation}
where $q$ stands for the momentum of the virtual $W$ boson. 
In general, the hadronic tensor can be expressed
in terms of five scalar functions $W_{1-5}(\nu,\, q^2)$ of the two 
independent Lorentz invariants, $\nu = q\cdot P/M_B$ and $q^2$ ($M_B$ and $P$
denote the mass and momentum of the $B$ meson, respectively),  
\begin{equation}
W_{\mu\nu} = -g_{\mu\nu}W_1 + \frac{P_{\mu}P_{\nu}}{M_B^2} W_2 
 -i\varepsilon_{\mu\nu\alpha\beta} \frac{P^{\alpha}q^{\beta}}{M_B^2}W_3
         + \frac{q_{\mu}q_{\nu}}{M_B^2} W_4 
 + \frac{P_{\mu}q_{\nu}+q_{\mu}P_{\nu}}{M_B^2} W_5 \, .
\label{eq:exp2}
\end{equation}
The theoretical task is to calculate the five structure functions.

Let's look into the expression (\ref{eq:comm2}) for the hadronic tensor. 
The current commutator in Eq.~(\ref{eq:comm2}) is causal,
$[j_{\mu}(y),j^{\dagger}_{\nu}(0)] = 0$ for $y^2<0$. Moreover,
$e^{iq\cdot y}$ in Eq.~(\ref{eq:comm2}) oscillates rapidly, averaging out
contributions to the integral except in the domain $y^2\leq 1/q^2$.  
For inclusive semileptonic $B$ decays,
the momentum transfer squared lies in the range $M_\ell^2\leq q^2
\leq (M_B-M_{X_{\rm min}})^2$, where $M_\ell$ is the charged lepton mass and
$M_{X_{\rm min}}$ is the minimal hadronic 
invariant mass in the final state, which is the $D$ meson (pion) mass
for the $b\to c$ ($b\to u$) transition.  
Because of the heaviness of the $B$ meson, the decays occur mostly at large 
momentum transfer. Therefore, the combined consideration of causality, less 
rapid oscillations and kinematics leads us to realize that the hadronic 
tensor is dominated by the space-time near the light cone $y^2\to 0$.

Another argument for the light-cone dominance is as follows. One can express
the commutator of two currents in terms of the bilocal operator:
\begin{equation}
\left[ j_{\mu}(y),j_{\nu}^{\dagger}(0) \right]
 = 2(S_{\mu\alpha\nu\beta} -i\varepsilon_{\mu\alpha\nu\beta})
  \left[ \partial^{\alpha}\Delta_q (y) \right] \bar{b}(0)
    \gamma^{\beta}{\cal P}exp[ig_s\int_y^0 dz^\mu A_\mu (z)]b(y)\, ,
\label{eq:domin3}
\end{equation}
where $S_{\mu\alpha\nu\beta} = g_{\mu\alpha}g_{\nu\beta} + g_{\mu\beta}
 g_{\nu\alpha} - g_{\mu\nu}g_{\alpha\beta}$ and $\Delta_q (y)$ is the
Pauli-Jordan function.
${\cal P}$ denotes path ordering. The coefficient function of the bilocal
operator in Eq.~(\ref{eq:domin3}) is finite except on the light cone
$y^2=0$ where it is singular. Therefore, the light-cone singularity also 
implies that the dominant contribution to the
hadronic tensor comes from the space-time separations in the neighborhood of
the light cone.

The light-cone dominance justifies the light-cone expansion in matrix elements
of increasing twist. The light-cone expansion provides a formal and 
powerful way of organizing the
nonperturbative QCD effects and singling out the leading term. 
The leading nonperturbative QCD contribution to
inclusive semileptonic decays of $B$ mesons resides in the distribution
function \cite{eur}
\begin{equation}
f(\xi) = \frac{1}{4\pi}\int \frac{d(y\cdot P)}{y\cdot P}e^{i\xi y\cdot P}
\langle B|\bar{b}(0)y\!\!\!/{\cal P}exp[ig_s\int_y^0 dz^\mu A_\mu (z)]b(y)
|B\rangle |_{y^2=0}\, .
\label{eq:distr3}
\end{equation}
The five structure functions, {\it a priori} independent,
are then related to a single distribution function in leading twist
approximation \cite{jp1,jin1,jp2,jin2,eur}:
\begin{eqnarray}
W_1 & = & 2[f(\xi_+) + f(\xi_-)]\, ,\label{eq:w1}\\
W_2 & = & \frac{8}{\xi_+ -\xi_-}[\xi_+f(\xi_+)-\xi_-f(\xi_-)]\,,\label{eq:w2}\\
W_3 & = & -\frac{4}{\xi_+-\xi_-} [f(\xi_+) -f(\xi_-)]\, ,\label{eq:w3}\\
W_4 & = & 0\, ,\label{eq:w4}\\
W_5 & = & W_3\, ,\label{eq:w5}
\end{eqnarray}
where $\xi_\pm = (\nu\pm\sqrt{\nu^2-q^2+m_q^2})/M_B$\, .

It has been shown \cite{eur} that
inclusive radiative decays of $B$ mesons $B\to X_s\gamma$ are also dominated
by light-cone dynamics. It turns out that the distribution function is 
universal: The same
distribution function encodes the leading nonperturbative QCD contribution to
inclusive radiative decays of $B$ mesons. The universality originates from
the fact that the primary object of analysis in long-distance effects is the
same bilocal operator matrix element dictated by light-cone dynamics.
In leading twist approximation, the photon energy spectrum is then given 
by \cite{eur}
\begin{equation}
\frac{d\Gamma(B\to X_s\gamma)}{dE_\gamma}= \frac{G_F^2\alpha m_b^2}{2\pi^4 M_B}
|V_{tb}V_{ts}^\ast|^2 |C^{(0)}_7(M_W)|^2 
E_\gamma^3f\Bigg (\frac{2E_\gamma}{M_B}\Bigg ).
\label{eq:spectrum}
\end{equation}
  
\section{Properties of the Distribution Function}
The $b$-quark distribution function $f(\xi)$ for the $B$ meson is a key 
object. Like
the well-known parton distribution functions for the nucleon in deep inelastic
lepton-nucleon scattering, the knowledge of the $b$-quark distribution function
would help us greatly in understanding the nature of confinement and the 
structure of the $B$ meson. I will survey our present knowledge of the
$b$-quark distribution function and look forward to likely advances in 
the future.

The distribution function is gauge invariant. It obeys positivity and the 
support of it is 
$0\leq\xi\leq 1$. The distribution function is exactly normalized to unity
\begin{equation}
\int_0^1 d\xi\, f(\xi) = 1 .
\label{eq:norm}
\end{equation}
This normalization does not get renormalized as a consequence of $b$-flavored
quantum number conservation. The distribution function contains the free quark
decay as a limiting case with $f(\xi) = \delta(\xi-m_b/M_B)$.
The distribution function $f(\xi)$ has a simple physical interpretation:
It is the probability of finding a $b$-quark with momentum $\xi P$ inside the
$B$ meson with momentum $P$.

In addition, the mean $\mu$ and the variance
$\sigma^2$ of the distribution function were deduced \cite{eur} 
using the techniques of the operator product expansion and the heavy quark 
effective theory (HQET):
\begin{equation}
\mu\equiv \int_0^1 d\xi\, \xi f(\xi)= \frac{m_b}{M_B}
\left( 1+\frac{5E_b}{3}\right) ,
\label{eq:mean}
\end{equation}
\begin{equation}
\sigma^2\equiv \int_0^1 d\xi\, (\xi-\mu)^2 f(\xi)= 
\left(\frac{m_b}{M_B}\right)^2
\left[\frac{2K_b}{3}-\left(\frac{5E_b}{3}\right)^2\right] ,
\label{eq:variance}
\end{equation}
where $E_b = K_b+G_b$ and $K_b$ and $G_b$ are the dimensionless HQET 
parameters of order $(\Lambda_{\rm QCD}/m_b)^2$, which are often referred to 
by the alternate names
$\lambda_1 = -2m_b^2K_b$ and $\lambda_2 = -2m_b^2G_b/3$. The parameter
$\lambda_2$ can be extracted from the $B^\ast-B$ mass splitting:
$\lambda_2=(M^2_{B^\ast}-M^2_B)/4=0.12$ GeV$^2$. The parameter $\lambda_1$
suffers from large uncertainty. 

The mean value and variance of the distribution function characterize the 
location of the ``center of mass'' of the distribution function and the
square of its width, respectively. They specify the primary shape of the
distribution function. From Eqs.~(\ref{eq:mean}) and (\ref{eq:variance}) we
know that the distribution function is sharply peaked around
$\xi = \mu \approx m_b/M_B$ close to 1 and its width of order 
$\Lambda_{\rm QCD}/M_B$ is narrow, suggesting that the distribution function
is close to the delta function form in the free quark limit. Note that
the $b$-quark distribution function has different forms for different 
beauty hadrons.\cite{jin2}

Nonperturbative QCD methods such as lattice gauge theory, light-cone field
theory, 
and QCD sum rules could help determine further the form of the distribution 
function. 

The $b$-quark distribution function can also be extracted directly from 
experiment. The $B\to X_s\gamma$ photon energy spectrum \cite{eur}
and the $B\to X_q\ell\nu$ spectra $d\Gamma/d\xi_+$ \cite{new,prd} share
a common feature --- namely they offer the intrinsically most sensitive probe 
of long-distance
strong interactions because these spectra correspond to a discrete line 
solely on kinematic grounds in the 
absence of gluon bremsstrahlung and long-distance strong interactions. 
Indeed, our calculation based on the light-cone expansion shows that they are 
explicitly proportional to the 
nonperturbative distribution function. Therefore, the shapes of these spectra
directly reflect the inner long-distance dynamics of the reactions and
measurements of these spectra are idealy suited for direct extraction of the 
distribution function from experiment. 

The universality of the distribution function implies
great predictive power: Once the distribution function is measured from one 
process, it can be used to make predictions
in all other processes in a model-independent manner. 
 
\section{Applications}
I have described the QCD approach to inclusive
decays of heavy hadrons from light-cone expansion and heavy quark effective 
theory. Because of the large mass of the decaying hadron, light-cone
dynamics plays a governing role in inclusive decays of heavy hadrons.
The light-cone expansion allows a rigorous and systematic
ordering of nonperturbative QCD effects, and the identification of the leading
term --- the $b$-quark distribution function. Another large scale set by the
$b$-quark mass allows the construction of the heavy quark effective theory,
providing another complementary framework for organizing and parametrizing 
nonperturbative
QCD effects. The additional properties of the distribution function have
been learned exploiting the heavy quark effective theory.

This approach has been applied to calculate decay rates and distributions
in semileptonic or radiative inclusive decays of $B$ mesons.
Model-independent predictions have been made from
the known normalization of the distribution function or the cancellation of 
the distribution function in the ratio of the decay rates. Some calculations of
decay rates and distributions involve the modelling of the distribution 
function,
since the form of it has as yet not been completely determined. However, for 
sufficiently inclusive quantities such as the total semileptonic decay rates,
the results are nearly model-independent, since they are essentially only
sensitive to the mean value and variance of the distribution function,
which are known from the heavy quark effective theory. Direct extraction of
the distribution function from precision measurements would eliminate the
model dependence. 

A crucial observation is that both dynamic and kinematic effects of 
nonperturbative QCD must be taken into account.\cite{jin1,pl,cern} 
The latter results in the
extension of phase space from the quark level determined by the $b$-quark mass
to the hadron level determined by the $B$ meson mass, thereby increasing 
decay rates. The calculations in
the light-cone approach are able to include both dynamic and kinematic 
effects of nonperturbative QCD, and have shown that
the net effect of nonperturbative QCD enhances the total semileptonic rates
for $B\to X_c\ell\nu$ \cite{jin1} and $B\to X_u\ell\nu$ \cite{pl}, contrary
to the results \cite{bigi} obtained in the heavy quark expansion approach.
The heavy quark expansion approach \cite{chay} assumes quark-hadron duality, 
and cannot account for the rate due to the extension of phase space from the
quark level to the hadron level. The light-cone approach provides a way around
problems associated with the assumption of quark-hadron duality.
The rate calculations in the light-cone approach have quantitatively 
shown \cite{jin1,pl}
the importance of the inclusion of kinematic effects of nonperturbative QCD.

The interplay between nonperturbative and perturbative QCD effects has been 
accounted for in the light-cone approach, since confinement implies that free 
quarks are not asymptotic states of the theory and the separation of 
perturbative and nonperturbative effects cannot be done in a clear-cut way.

The following are some of the results.

(1) A new method for precise and model-independent determination of $|V_{ub}|$ 
has been proposed.\cite{new,prd} 
For inclusive charmless semileptonic decays of $B$ mesons $B\to X_u\ell\nu$,  
the light-cone expansion and $b$-flavored quantum number conservation lead
to the sum rule \cite{new}
\begin{equation}
S\equiv\int_0^1 d\xi_u\, \frac{1}{\xi_u^5}
\frac{d\Gamma}{d\xi_u}(B\to X_u\ell\nu) 
= |V_{ub}|^2\frac{G_F^2M_B^5}{192\pi^3}\, ,
\label{eq:sumrule}
\end{equation}
with the kinematic variable $\xi_u = (q^0+|\vec{q}|)/M_B$ in the $B$-meson
rest frame.
This sum rule receives no perturbative QCD correction, and avoids the
dominant hadronic uncertainty. 
The sum rule (\ref{eq:sumrule}) thus establishes a clean relationship 
between $|V_{ub}|$ and the observable $S$, allowing precise 
determination of $|V_{ub}|$. This determination of $|V_{ub}|$ is also
model-independent in the sense that the sum rule is independent of 
phenomenological models. Moreover, this method is not only exceptionally
clean theoretically, but also very efficient experimentally in background
suppression. Applying the kinematic cut $q^0 > M_B-M_D$ or $M_X < M_D$,
which leads to $\xi_u > 1-M_D/M_B$, one can discriminate between $b\to u$
signal and $b\to c$ background. The $\xi_u$ spectrum is unique. About $80\%$
of the spectrum satisfy $\xi_u > 1-M_D/M_B$.

(2) $|V_{ub}|$ has been determined \cite{pl} 
from the measured inclusive charmless semileptonic branching ratio of
beauty hadrons.
The determination of $|V_{ub}|$ from the measurement of the inclusive 
charmless semileptonic branching ratio in conjunction with the calculation of 
the total charmless semileptonic rate has larger theoretical uncertainty than
the proposed determination of $|V_{ub}|$
from the measurement of the observable $S$ in conjunction with the sum 
rule (\ref{eq:sumrule}),
since the calculations of perturbative
and nonperturbative QCD corrections to the total charmless semileptonic rate
suffer from much larger theoretical uncertainty \cite{pl} than 
the sum rule (\ref{eq:sumrule}), not to mention the fundamental uncertainty
due to the assumption of quark-hadron duality if the total semileptonic rate
is calculated in the heavy quark expansion approach.

(3) The $B\to X_u\ell\nu$ ($\ell = e$ or $\mu$) charged lepton energy 
spectrum \cite{jp2}, 
hadronic invariant mass spectrum \cite{jin3}, and lepton pair 
spectrum \cite{prd} have been analyzed in detail, resulting in the 
state-of-the-art descriptions of these spectra. It has been shown that 
the interplay between perturbative and nonperturbative QCD eliminates the
singularities of the parton-level perturbative spectra.  

(4) $|V_{cb}|$ has been determined \cite{jin1} from the measured inclusive 
semileptonic branching ratio of beauty hadrons in conjunction with the 
calculation of the 
total semileptonic rate. The charged lepton energy spectrum in inclusive
semileptonic decays of $B$ mesons $B\to X_c\ell\nu$ has been 
analyzed \cite{jp2} in detail, resulting in the state-of-the-art description
of the spectrum. The calculated spectrum was found \cite{jp2} 
to be in agreement with the experimental data, providing an experimental test 
of the validity of the light-cone approach.

(5) The total semileptonic decay rate of the $\Lambda_b$ baryon has been
calculated.\cite{jin2} The calculated semileptonic branching ratio for 
$\Lambda_b$ is consistent with the measurements.

(6) Nonperturbative QCD effects on the $B\to X_s\gamma$ photon energy spectrum
have been calculated.\cite{eur} The theoretically clean methods for the 
determinations of $|V_{ts}|$, $|V_{ts}/V_{ub}|$ and $|V_{ts}/V_{cb}|$ have 
been suggested.\cite{eur}

\section*{Acknowledgments}
I am indebted to Emmanuel Paschos for collaboration. I would like to thank 
Stanley Brodsky, Chueng-Ryong Ji and Bruce McKellar for useful discussions.
This work was supported by the Australian Research Council.

\end{document}